\documentclass{kluwer}
\usepackage{graphicx}
\begin{document}
\begin{article}
\begin{opening}
\title{Are there any Young Galaxies in our local universe?}            

\author{Daniel \surname{Kunth}\email{kunth@iap.fr}} 
\institute{Institut d'Astrophysique de Paris, 98bis Boulevard Arago, 75014 
Paris, France}                               
\author{G\"oran \surname{\"Ostlin}\email{ostlin@astro.su.se}}
\institute{Stockholm Observatory, 133 36 Saltsj\"obaden, Sweden}

\date{August 3, 2000 } 

\runningtitle{Census on the existence of Young Galaxies...}
\runningauthor{Daniel Kunth and G\"oran \"Ostlin}

\begin{abstract}

We find no evidence for the presence of genuinely young galaxies in our Local 
universe.
We argue that the metal deficiencies and the behavior of the N/O and C/O
ratios meet several plausible explanations involving the primary and
secondary nature of these elements and their dispersion and mixing by massive 
stars and intermediate mass stars. Colors are to be analysed using surface
photometry and comparing several model scenario since an instant burst model
unavoidably provides a lower limit to the age. Ultimately  colour-magnitude 
diagram  techniques will be the only direct test to be performed at least 
on a small number local starburst galaxies. Our present analysis provides
important warning for assessing the true age of distant galaxies.

\end{abstract}


\end{opening}
           
\section{Introduction}
It remains very important in the context of galaxy formation to establish 
beyond  doubt whether or not galaxy formation is a continuous process so as to 
expect 
genuinely young galaxies to be observed in our Local Universe. This issue 
bears with some difficulties for many reasons, as for example the semantic 
aspect of what  means ``young''. Here we shall adopt the view that a young 
galaxy is  a galaxy at its very first episode of star formation. 
Observationaly the
local dwarf starbursts discovered in the early 70s (Searle and Sargent, 1972) 
qualified for youth
because they appeared to be metal deficient, gas-rich, have 
very blue colors, and for many of them, small amounts of dust. If 
such galaxies were young one should  be able to detect the H{\sc i} clouds out 
of 
which they are formed. The first survey by Lo and Sargent (1979) concluded for 
the absence of any excess of gas rich dwarfs and that no H{\sc i} clouds without 
an 
optical counterpart had been found.
 At this moment, the H{\sc i} mass function is still ill-defined (Martin 
Zwaan, 1998) but really constraining would be an 
accurate determination below, say, $10{^7}$ 
$M_\odot$. 
 However the fact that  galaxies 
contain young blue massive stars does not necessarly imply that they are
 young. In fact, most blue compact dwarfs simply experience, at present,
a burst of star formation which is, by no means, the first of the kind. With 
the advent of modern detectors more than 90$\%$ reveal an underlying old 
stellar population (Kunth and 
\"Ostlin, 2000).

What about the rest?  At this point we wish
to argue that since most galaxies in the neighbourhood contain old stars, a
genuinely young object would be a quite sensational discovery, and have 
more far--reaching  implications than finding just another old galaxy. 
Therefore the burden of proof  has to be carried by the youth hypothesis.
It has to be shown beyond reasonable doubt (the limit set by existing 
technology) that an old population is absent, bearing in mind that it is much 
easier to detect young stellar populations than old ones. We shall examine 
how one could rely on metallicity determination and on colors to establish 
that a galaxy is young. Ultimately, colour-magnitude diagram  (CMD) techniques 
will be the only direct test to be performed, keeping in mind however that it, 
up to now, has been applicable only to a very limited number of blue compact
galaxies.

\section{Metals}

 H{\sc ii} region 
abundance  analysis can  provide abundances of heavy elements such as O, C, 
and N that are of extreme importance in chemical evolution models. While 
metallicity is an important tool for studying how a galaxy chemically 
evolves, 
it is by no means straightforward to infer anything about the age (Kunth and 
\"Ostlin, 2000).
This issue has also been  discussed by M. Edmunds in this conference.

Izotov and Thuan (1999),  hereafter IT99, have observed a large set of
blue compact galaxies among the most O/H deficient. They find positive 
correlations between C/O and N/O
with O/H, but for 12+log(O/H) $ \le 7.6$,  
C/O and N/O remain constant and independent of O/H. Moreover, at these low
oxygen abundances their N/O and C/O show almost no scatter, contrary to 
 higher O/H. They  conclude that
galaxies with such low abundances are genuinely young (less than 40 Myr old), 
hence making their first generation of stars while
all galaxies with 7.6 $\le$ 12+log(O/H) $\le$ 8.2 have ages significantly 
smaller than 1 Gyr. The basis of such a statement is that C, N and O should be 
produced by the ongoing burst as primary elements in  galaxies with 
12+log(O/H) $ \le 7.6$, meaning that this observed burst must be the first 
one.  The absence of scatter is  interpreted as evidence that intermediate
mass stars (IMS) have not yet contributed to the N enrichment (because this 
would introduce time delays, hence a scatter). For galaxies with 12+log(O/H) 
$ > 7.6$ N is secondary and N/O increases with time. 

 
Admitedly the N/O diagram shows very little scatter at 12+log(O/H) $ \le 7.6$, 
which is somewhat surprising from  an observational point of view (N lines are 
in principle difficult to detect at such low abundances and there are many 
corrections involved in order to get both a correct O and N value). But even 
at face value the absence of a scatter does not necessarily have anything to 
do with youth. 

Indeed chemical evolution models predict N/O to increase rapidly 
and level out around $\sim -1.5$ (e.g. Olofsson 1995). Subsequent starbursts 
will increase O/H and
introduce wiggles (i.e. a scatter) in N/O since N production is delayed with
respect to O. However, if star formation is continuous (at least prior to the
present burst)  no scatter in N/O is expected. The same is true if bursts
are separated by long enough times that N can catch up with O (Pilyugin 1999).
The selection criteria has, on purpose (since the common belief has been that a 
young galaxy 
should look similar to IZw18), picked out actively star forming galaxies, 
and will miss dormant post burst objects where N/O may have  decreased. 
Connected
to this is the question of the mixing timescale of fresh metals: If IT99
are correct (age less than 40 Myr and O/H a function of time), this must be 
close to instantaneous and similar to the timescale for 
massive star evolution. However the relative CNO yields are a sensitive function
of mass, hence time, after the onset of the burst. Thus, if mixing is as fast 
as the youth scenario needs, we should expect to observe a scatter at low age,
and IT99 do not. If the mixing timescale is longer, of the order of 100 Myr,
the scatter could disappear before it becomes  observable. 

Henry et al. (2000) and Edmunds (this conference) advocate that IMS can
account for the bulk of primary N hence in this case we observe 
the average N/O obtained after burst completion. Timescale for 
complete dispersion and mixing needs to be larger than a Gyr (see also Piluygin 
1999). The increase of N/O and C/O at higher metallicity can be explained by a 
metallicity
dependent yield, perhaps with contributions from WR stars.

Another illustration that O/H in itself is by no means an indication of youth
is that SBS 0335-052 (one of the most  O/H deficient galaxies) has a similar 
O/H abundance in 
two knots that are completely unrelated causually (Lipovetsky et al., 1999).
 In IZw18 both H{\sc i} and H{\sc ii} metallicity is now known to be similar and 
show no 
gradient indicating that dispersion and mixing  must
have operated after one or several star formation episodes, excluding the 
possibility that the observed metals result from the present burst (van Zee 
et al., 1998; Legrand et al., 2000, Legrand, 2000). 
Finally we find it rather 
unplausible that new SN-released metals could be incorporated into the ionised 
gas 
during the short lifetime of the burst. Rather they are more likely to remain 
for a long time in a hot phase before they cool down and mix (Pantelaki and 
Clayton, 1987; Tenorio-Tagle, 1996).

\section{Colors}

Several studies have attempted to reveal an old stellar population in the
outskirts of blue compact starbursts using surface photometry. It is easy to 
show that in the red or near infrared, old stars should be easier to detect 
than in the blue, where massive stars dominate. 
Old stars have to dominate the young stellar population  in order
to produce an observable impact on the colours. 
When using optical bandpasses ($U,B,V,R$) only, a single stellar 
population 
(SSP) of 1 Gyr must  account for at least 90 \% of the mass to be detectable 
against a 10 Myr SSP\footnote{These estimates made use of the PEGASE.2 spectral 
synthesis code (Fioc and Rocca-Volmerange 1999).}. Inclusion of near-IR 
passbands reduces this figure to 
$\sim$50 \%. 
For a 10 Gyr old SSP, these figures rise to $\sim$95\% and $\sim$75\% 
respectively.
Observed colors have to be compared with predictions from spectral synthesis 
models. We emphasize here that many authors bias their discussion when using 
only one model scenario -- presented to be the simplest -- such as the instant 
burst model (= SSP). In fact this model 
will unavoidably provide a lower limit to the age as we show in Fig.  Using only
a SSP model to estimate the age is not only the ``simplest'' -- it is 
also an oversimplification.

Another problem in interpreting colours is the possible contribution from 
ionised 
gas, both in the form of line - and continous emission.  Although some spectral 
synthesis models include gas emission, these assume that UV-photons are 
produced locally, whereas in the halo of a BCG, they are likely to leak out from 
the burst region. 
Gas emission makes some colours bluer (e.g. $V-I_{\rm{\sc c}}$), 
others 
redder (e.g. $B-R_{\rm{\sc c}}$), but the effect also depends on gas metallicity 
and the 
hardness of the ionising spectrum.  By using combined colour indices, 
which tend to either increase or decrease the estimated ages, the problem 
can be circumvined somewhat. Using deep spectra  helps to estimate the 
emission-line contribution along the 
slit  (e.g. Papaderos et al. 1998), however this technique is limited if the 
relative 
contribution 
from ionised gas varies with position angle. Indeed, there are indications that 
dwarf galaxies have extended gas emission along the \emph{minor} axes, in line 
with 
simple models of winds and outflows.

In Fig. 1 we compare published halo colours for two young galaxy candidates:
Tololo~65 and SBS~0335-052 with the predictions from spectral synthesis 
models (PEGASE.2 by Fioc and Rocca-Volmerange, 1999). For Tololo~65 the 
agreement  
is good when comparing models with extended (i.e. not SSP) star formation 
history. 
For SBS~0335-052, $U-B$ gives a lower estimate, still the data can be fit with 
an 
age of several Gyr. There is nothing from the photometrical point of view that 
\emph{requires} these galaxies to be young.
Similarly, for IZw18, the surface photometry consistently gives a lower 
limit of 1 Gyr (\"Ostlin, this conference) in agreement with the results 
from CMD analysis (Aloisi et al. 1999; \"Ostlin 2000), although there may be a 
serious 
contribution from ionised gas in the halo (Izotov 2000, private correspondance).

\begin{figure}
\hspace{-0.575cm}
\includegraphics[width=15.5pc]{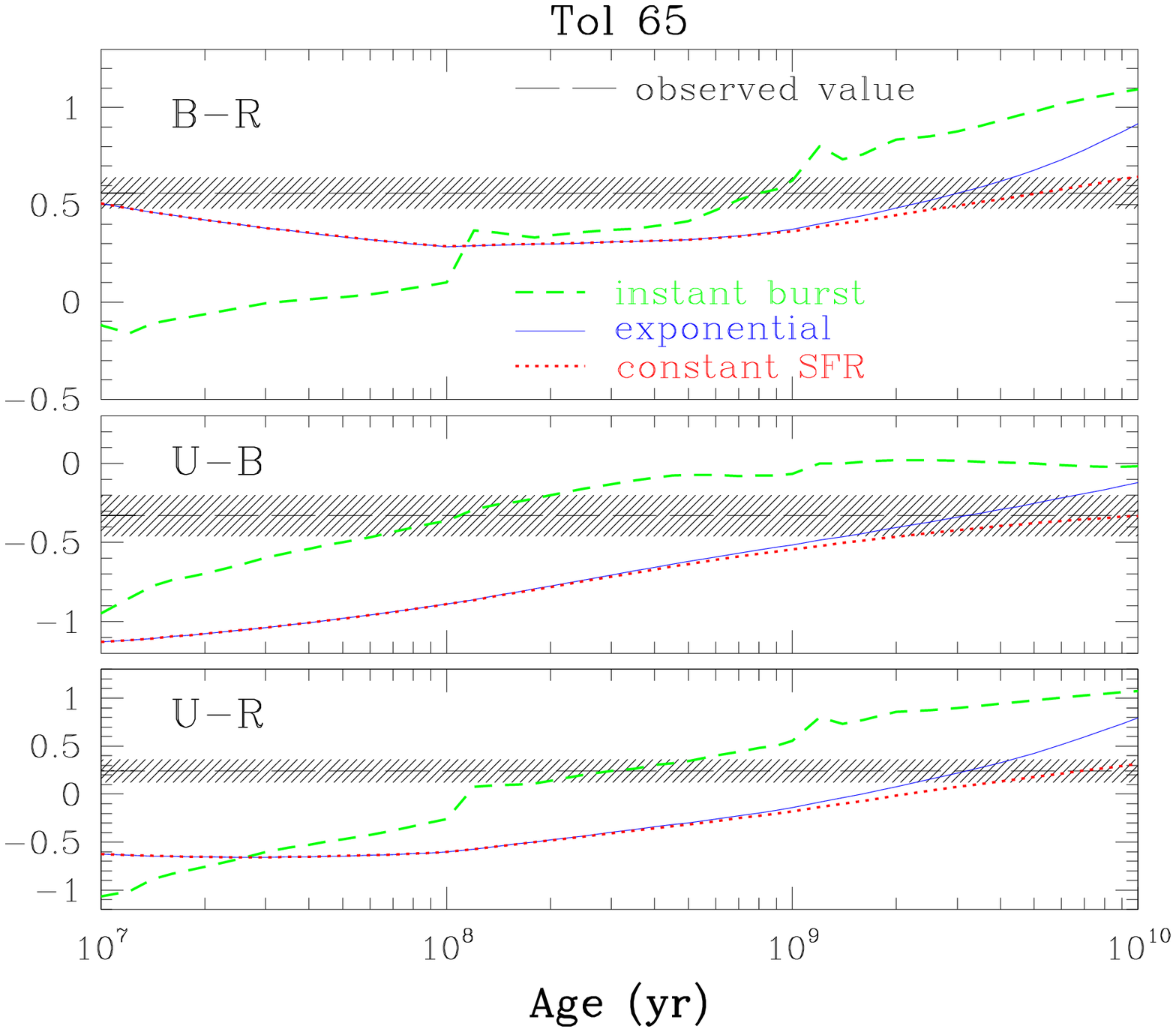}
\hspace{-0.45cm}
\includegraphics[width=15.5pc]{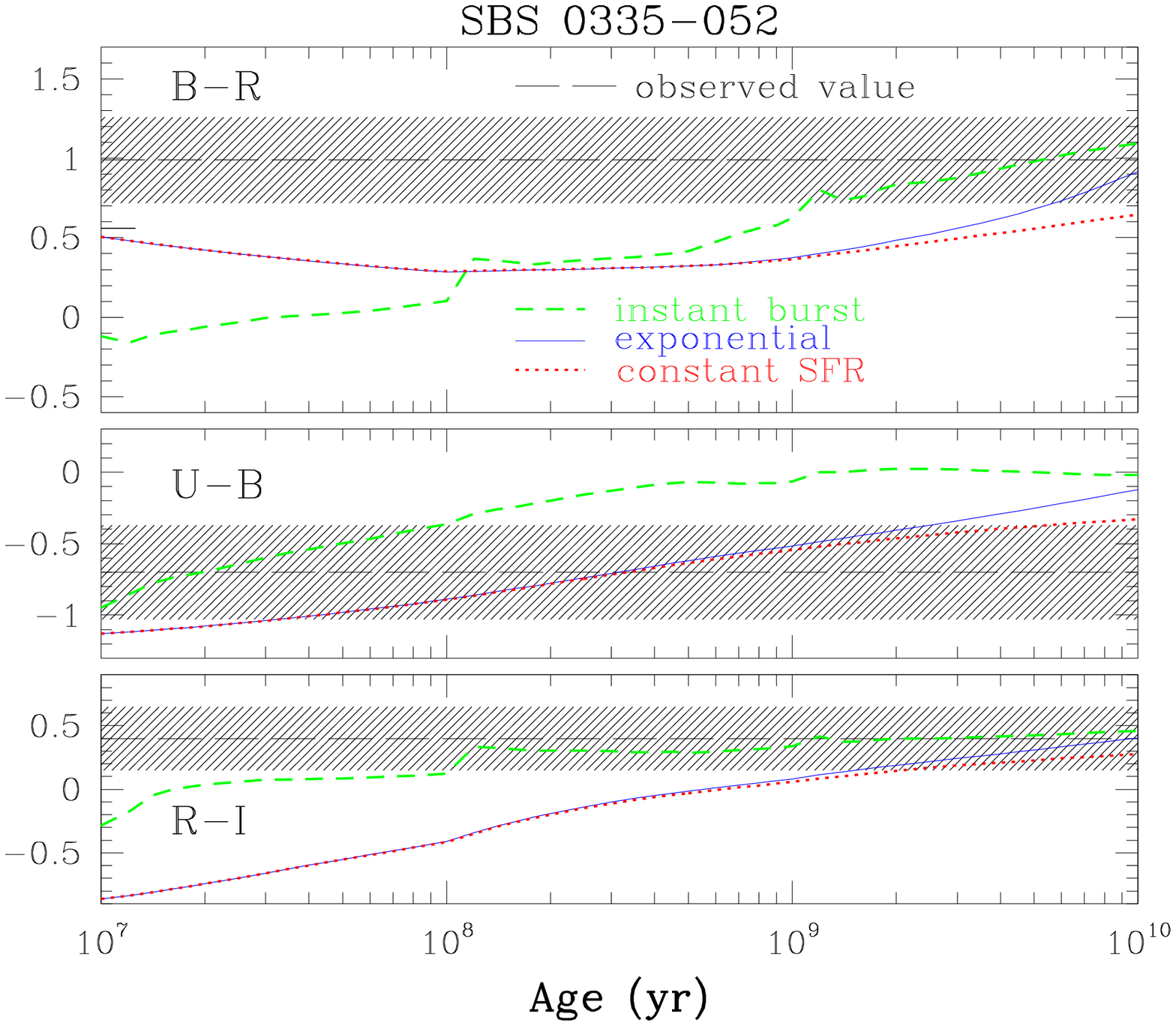}
\caption{The broadband colours in the haloes of Tololo 65 (left) and
SBS~0335--052 (right)  compared with the  model predictions from PEGASE.2 for 
three 
different star formation histories. Additional used parameters: $Z=0.004$ and 
standard 
Salpeter IMF with mass range 0.1 to $120 M_\odot$.
The shaded region gives the quoted 
observational uncertainty. 
It is evident that the instant burst approximation gives a lower limit to the 
age. 
The best fit is produced assuming an exponentially decaying star formation rate 
(e-folding
time 3 Gyr) which yields an age of several Gyr in both cases. 
The photometric 
data is 
from Papaderos et al. (1999) for Tololo 65 and from Fig. 5 
in Papaderos et al. (1998) for SBS 0335-052.}   
\end{figure}

\section{Conclusion}

No conclusive evidence remains that metallicity and colours are unanbiguous 
tracers of youth in any galaxy of our local Universe. We have outlined the
danger in oversimplifying the analysis of metal ratios and colour indices
using only a SSP scenario. We conclude that care should be taken before
assessing an age to distant galaxies.
 This show that the only way out, in our local
universe, is to study very nearby galaxies with CMD techniques.

\begin{acknowledgements}

We thank N. Prantzos and K. Olofsson for stimulating discussions on chemical 
evolution,
and Y. Izotov for interesting discussions on the nature of IZw18.

\end{acknowledgements}

\end{article}

\begin{thebibliography}{}

\bibitem[\protect\citeauthoryear{Aloisi et al.}{1999}]{aloisi}
Aloisi, A., Tosi, M. and Greggio, L.: 2000, {\it AJ}  {\bf 118}, 302

\bibitem[\protect\citeauthoryear{Fioc and Rocca-Volmerange}{1999}]{pegase}
Fioc, M. and Rocca-Volmerange, B.: 1999, astro-ph/9912179

\bibitem[\protect\citeauthoryear{Henry et al.}{2000}]{henry}
Henry, R.B.C., Edmunds, M.G. and K\"oppen, J.: 2000, {\it ApJ} in press 
(astro-ph/0004299) 

\bibitem[\protect\citeauthoryear{Izotov and Thuan}{1999}]{it}
Izotov, Y.I. and Thuan, T.X.: 1999, {\it ApJ} {\bf 511}, 639

\bibitem[\protect\citeauthoryear{Kunth and \"Ostlin}{2000}]{kunthostlin}
Kunth, D. and \"Ostlin, G.: 2000,  {\it A\&AR}  {\bf 10}, 1

\bibitem[\protect\citeauthoryear{}{}]{}
Legrand F.: 2000 {\it A\& A} {\bf 354}, 504

\bibitem[\protect\citeauthoryear{}{}]{}
Legrand F., Kunth D., Roy J.-R., Mas-Hesse J.M. and Walsh J.R.: 2000, {\it A\& 
A}
 {\bf 355}, 891

\bibitem[\protect\citeauthoryear{}{}]{}
Lipovetsky, V.A., Chaffee, F.H., Izotov, Y.I., Foltz, C.B., Kniazev, A.Y. and
Hopp, U.: 1999, {\it ApJ} {\bf 519}, 177

\bibitem[\protect\citeauthoryear{}{}]{}
Lo K.Y., Sargent W.L.W.: 1979, {\it ApJ} {\bf227}, 756

\bibitem[\protect\citeauthoryear{\"Ostlin}{2000}]{ostlin}
\"Ostlin, G.: 2000, {\it ApJ}  {\bf 235}, L99

\bibitem[\protect\citeauthoryear{}{}]{}
Pantelaki I., Clayton D.D., 1987, in ``Starbursts and galaxy evolution'', Eds. 
Thuan T.X., Montmerle T., Tran Thanh Van J., \'Editions Fronti\`eres, p. 145

\bibitem[\protect\citeauthoryear{Papaderos et al.}{1998}]{p98}
Papaderos, P., Izotov, Y.I., Fricke, K.J., Thuan, T. X. and
 Guseva, N.G.: 1998, {\it A\&A} {\bf 338}, 43

\bibitem[\protect\citeauthoryear{Papaderos et al.}{1999}]{p99}
Papaderos, P., Fricke, K.J., Thuan, T.X., Izotov, Y.I. and
 Nicklas, H.:  1999 {\it A\& A} {\bf 352}, 57

          
\bibitem[\protect\citeauthoryear{}{}]{}
 Searle, L. and Sargent, W.L.W.: 1972, {\it ApJ} {\bf 173}, 25

\bibitem[\protect\citeauthoryear{}{}]{}
Tenorio-Tagle G.: 1996, {\it AJ} {\bf 111}, 1641 

\bibitem[\protect\citeauthoryear{}{}]{}
van Zee L., Haynes M.P., Salzer J.J., Broeils A.H.: 1998, {\it AJ} {\bf 115} ,
 1000

\bibitem[\protect\citeauthoryear{}{}]{}
Zwaan M. 1998, In "Dwarf Galaxies and Cosmology" Proceedings of
The Astrophysics Session of the XXXIIIth Rencontres de Moriond. eds: T.X.
Thuan, Balkowsky C., Cayatte V and J.Tran Thanh Van. Ed. Frontieres, p.49

\end{thebibliography}
\end{document}